\documentclass[a4paper,12pt]{article}
\usepackage{jheppub,esint,shuffle,psfrag}
\usepackage[utf8]{inputenc}

\usepackage{epsfig,amssymb,amsmath,psfrag,subfigure,rotate,color,wasysym,xcolor}
\usepackage[bbgreekl]{mathbbol}

\allowdisplaybreaks
\newcommand{\insertfig}[2]{\includegraphics[width=#1cm]{#2}}

\DeclareSymbolFontAlphabet{\mathbbm}{bbold}
\DeclareSymbolFontAlphabet{\mathbb}{AMSb}%

\def\XXint#1#2#3{{\setbox0=\hbox{$#1{#2#3}{\int}$ }
\vcenter{\hbox{$#2#3$ }}\kern-.6\wd0}}

\def \be  {\begin{equation}}
\def \ee  {\end{equation}}
\def \ba  {\begin{eqnarray}}
\def \ea  {\end{eqnarray}}
\def \baa {\begin{eqnarray*}}
\def \eaa {\end{eqnarray*}}
\def \lab #1 {\label{#1}}

\newcommand\re[1]{(\ref{#1})}
\def\d{\hbox{{d}\kern-.20em\hbox{l}}}
\def \qqquad {\qquad\quad}
\def \qqqquad {\qquad\qquad}
\def \matrix #1 {\left(\begin{array}{cc} #1 \end{array}\right)}

\def \tr {\mathop{\rm tr}\nolimits}

\def \e  {\mathop{\rm e}\nolimits}
\newcommand\lr[1]{{\left({#1}\right)}}

\newcommand \vev [1] {\langle{#1}\rangle}

\newcommand \ket [1] {|{#1}\rangle}
\newcommand \bra [1] {\langle {#1}|}

\def\1{\hbox{{1}\kern-.25em\hbox{l}}}

\newcommand{\ft}[2]{{\textstyle\frac{#1}{#2}}}
\makeatletter

\input pdf-trans
\newbox\qbox
\def\usecolor#1{\csname\string\color@#1\endcsname\space}
\newcommand\bordercolor[1]{\colsplit{1}{#1}}
\newcommand\fillcolor[1]{\colsplit{0}{#1}}
\newcommand\outline[1]{\leavevmode%
  \def\maltext{#1}%
  \setbox\qbox=\hbox{\maltext}%
  \boxgs{Q q 2 Tr \thickness\space w \fillcol\space \bordercol\space}{}%
  \copy\qbox%
}
\makeatother
\newcommand\colsplit[2]{\colorlet{tmpcolor}{#2}\edef\tmp{\usecolor{tmpcolor}}%
  \def\tmpB{}\expandafter\colsplithelp\tmp\relax%
  \ifnum0=#1\relax\edef\fillcol{\tmpB}\else\edef\bordercol{\tmpC}\fi}
\def\colsplithelp#1#2 #3\relax{%
  \edef\tmpB{\tmpB#1#2 }%
  \ifnum `#1>`9\relax\def\tmpC{#3}\else\colsplithelp#3\relax\fi
}
\bordercolor{black}
\fillcolor{white}
\def\thickness{.3}

\def\1{\mathbbm{1}}

\title{Exact null octagon}
 
\author[a,b]{A.V.~Belitsky}
\author [b]{and G.P.~Korchemsky}
 \affiliation[a] {Department of Physics, Arizona State University, 
 Tempe, AZ 85287-1504, USA}

 \affiliation[b] {Institut de Physique Th\'eorique\footnote{Unit\'e Mixte de Recherche 3681 du CNRS}, Universit\'e Paris Saclay, CNRS, CEA, 91191 Gif-sur-Yvette} 
 
\preprint{  \parbox[t]{28mm}{IPhT--T19/097}}

 \abstract
{
We consider the so-called simplest correlation function of four infinitely heavy half-BPS operators in planar $\mathcal N=4$ SYM  in the limit when the operators are 
light-like separated in a sequential manner. We find a closed-form expression for the correlation function in this limit as a function of the 't Hooft coupling and residual 
cross ratios. Our analysis heavily relies on the factorization of the correlation function into the product of null octagons and on the recently established determinant 
representation for the latter. We show that the null octagon is given by a Fredholm determinant of a certain integral operator which has a striking similarity to those 
previously encountered in the study of two-point correlation functions in exactly solvable models at finite temperature and of level spacing distributions in random 
matrices. This allows us to compute the null octagon exactly by employing a method of differential equations.
 }

\begin{document}

\maketitle
\flushbottom
\setcounter{footnote} 0

%\newpage

%%%%%%%%%%%%%%%%%%%%%%%%%%%%%%%%%%%%%%%%%%%%%%%%%%%%%%%%%%%%%%%%%%%%%%%%%%%
\section{Introduction and summary of the results}
%%%%%%%%%%%%%%%%%%%%%%%%%%%%%%%%%%%%%%%%%%%%%%%%%%%%%%%%%%%%%%%%%%%%%%%%%%%

Recently, an important progress has been achieved in understanding the properties of four-point correlation functions of operators in four-dimensional maximally supersymmetric Yang-Mills 
theory ($\mathcal N=4$ SYM) based on integrability approach. According to the gauge/string duality \cite{Maldacena:1997re}, these correlation functions are dual to scattering amplitudes of 
four closed string states propagating on the AdS$_5\times$S$^5$ background. A novel approach to computing such amplitudes that makes full use of integrability of the underlying 
two-dimensional string sigma model was put forward in \cite{Basso:2015zoa}.  It is based on a tessellation of  the string world-sheet in a particular fashion through off-shell open string 
transitions known as hexagons \cite{Basso:2015zoa,Fleury:2016ykk,Eden:2016xvg,Bajnok:2017mdf}. 

This approach allows one to describe four-point correlation functions of any operators in planar $\mathcal N=4$ SYM for arbitrary 't Hooft coupling. However its application to computing correlation functions of light operators at weak coupling, where a plethora of data is available  from more traditional approaches like perturbative CFT calculations and analytic bootstrap, is hampered by significant technical difficulties  even to lowest order in coupling. On the other hand, in the limit of very heavy operators, where the conventional methods are not efficient, the hexagonalization approach is advantageous. In particular,  under a judicious choice of four infinitely heavy half-BPS operators, their four-point correlation function factorizes into a double 
copy of open string partition function, which was dubbed the octagon \cite{Coronado:2018ypq}. The hexagonalization is then operationally effective in working out an explicit form 
of the latter \cite{Coronado:2018ypq,Coronado:2018cxj}.

The `simplest' correlation function that can be described in terms of the octagon takes the form
\begin{align} \label{G4}
G_4 
= 
\vev{\mathcal{O}_1(x_1) \mathcal{O}_2(x_2) \mathcal{O}_3(x_3) \mathcal{O}_1(x_4)}
=
{ \mathcal{G}(z,\bar z) \over (x_{12}^2 x_{13}^2 x_{24}^2x_{34}^2)^{K/2}} \,,
\end{align}
where the half-BPS operators are built out of two complex scalars $Z$ and $X$ and their complex conjugate partners, $\mathcal{O}_1 =  \tr(Z^{K/2} \bar X^{K/2}) + \text{permutations}$, $\mathcal{O}_2 = \tr(X^K)$ and $\mathcal{O}_3 = \tr(\bar Z^K)$. They have the same scaling dimension $K\gg 1$ which is assumed to be even.  
Due to a particular choice of the half-BPS operators, at tree level (for zero coupling) the correlation function $G_4$ in the planar limit is given by the product of free scalar propagators connecting adjacent operators. 

\begin{figure}[t]
\begin{center}
\mbox{
\begin{picture}(0,180)(260,0)
\put(1,-260){\insertfig{20}{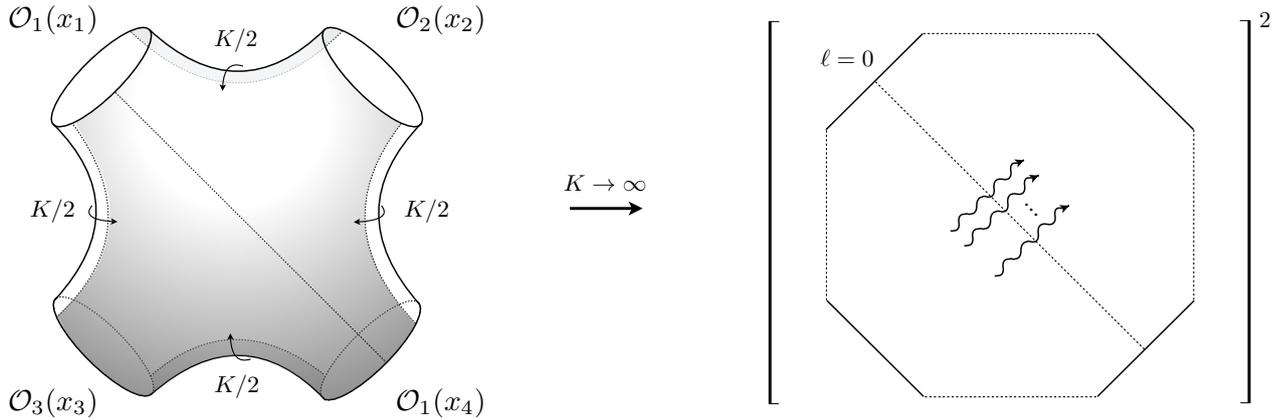}}
\end{picture}
}
\end{center}
\caption{\label{CorrelationOctagonFig} Planar four-point correlation function \re{G4} drawn on a sphere with four punctures (left panel). For $K\to \infty$ it
factorizes into the product of two identical octagons (right panel). The latter in turn are tessellated in terms of hexagons connected along a zero-length bridge (dashed line). The octagon is given by an infinite sum over excitations propagating on the two-dimensional world-sheet (shown by wiggly lines). }
\end{figure}

Quantum corrections to $G_4$ are described by the function  $\mathcal{G}$. It depends on the 't Hooft coupling $g^2 = g_{\rm\scriptscriptstyle YM}^2 N_c /(4 \pi)^2$, the length of the operators $K$ and kinematical variables $z$ and $\bar z$ related to  
 the two cross ratios 
\begin{align}\label{cross}
 {x_{12}^2 x_{34}^2\over x_{13}^2 x_{24}^2}=z \bar z\,,\qqqquad
 {x_{23}^2 x_{41}^2\over x_{13}^2 x_{24}^2}=(1-z)(1-\bar z)\,.
\end{align}
In the planar limit, $G_4$ receives contributions from Feynman diagrams that have the topology of a sphere with four punctures standing for operator insertions as shown in Fig.\ \ref{CorrelationOctagonFig} (left panel). Cutting the sphere into two faces along $K/2$  scalar propagators connecting the adjacent operators, one finds that at large $K$ the interactions between the two faces is suppressed for arbitrary coupling. Then, in the  limit $K \to \infty$, the interaction between the two faces is 
turned off and, as a consequence, the correlation function  factorizes into a 
product of two identical octagons \cite{Coronado:2018ypq,Coronado:2018cxj}~\footnote{In general, the octagon depends on some additional parameter, a bridge length $\ell$. The correlation function \re{G4} is related to the `simplest' octagon with $\ell=0$.}, as demonstrated in the right panel of Fig.\ \ref{CorrelationOctagonFig},
\begin{align}\label{G=O2}
\mathcal{G} (z,\bar z)  \stackrel{K \to \infty}{=} [\mathbb{O} (z,\bar z)]^2
\, .
\end{align}
At weak coupling, this relation holds for finite $K$ up to corrections of order $O(g^{K+2})$. 

%%%%%%%%%%%%%%%%%%%%%%%%%%%%%%%%%%%%%%%%%%%%%%%%%%%%%%%%%%%%%%%%%%%%%
%            Figure
%%%%%%%%%%%%%%%%%%%%%%%%%%%%%%%%%%%%%%%%%%%%%%%%%%%%%%%%%%%%%%%%%%%%%
%%%%%%%%%%%%%%%%%%%%%%%%%%%%%%%%%%%%%%%%%%%%%%%%%%%%%%%%%%%%%%%%%%%%%

The explicit expression for the octagon $\mathbb{O} (z,\bar z)$ at weak coupling was derived in  Refs.~\cite{Coronado:2018ypq,Coronado:2018cxj}. It takes the form
\begin{align}
\label{OctagonLadders}
\mathbb{O}(z,\bar z)  = 1 - g^2 f_1 + 2 g^4 f_2 - 6 g^6 f_3   + g^8 \lr{ 20f_4 -\frac13 f_2^2 + f_1 f_3} + O(g^{10})
\, ,
\end{align}
where the coefficient in front of $g^{2n}$ is given by  a multilinear combination $\sum c_{j_1,\dots,j_k} f_{i_1}\dots f_{i_k}$ (with $i_1+\dots+i_k=n$) of the so-called ladder integrals  
\begin{align} \label{data1}
f_L = {(1-z)(1-\bar z)\over z-\bar z}\sum_{m=0}^L {(-1)^m (2L-m)! \over L! (L-m)! m!} \log^m(z\bar z) \lr{\text{Li}_{2L-m}(z) - \text{Li}_{2L-m}(\bar z) }
\, .
\end{align}
As was shown in Refs.~\cite{Coronado:2018ypq,Coronado:2018cxj}, the expansion coefficients $c_{j_1,\dots,j_k}$ can be determined to all loops by requiring $\mathbb{O}(z,\bar z)$ to have an appropriate asymptotic behavior in two different OPE limits: (i) at short distances $z,\bar z\to 1$ and (ii) in the light-like limit $x_{12}^2,x_{13}^2,x_{24}^2,x_{34}^2\to 0$ where the four operators in \re{G4} are located at the vertices of a null rectangle.

It is instructive to compare the `simplest' correlation function \re{G=O2} and \re{OctagonLadders} with a generic four-point correlation function of half-BPS operators of an arbitrary length $K$. The latter function has been constructed in Refs.~\cite{Eden:2012tu,Chicherin:2015edu} by exploiting its properties in the two OPE limits alluded to above. It is given by a multi-linear combination of conformal integrals similar to \re{OctagonLadders}. The important difference is however that the ladder integrals do not furnish the complete basis of functions in this case. Starting from three loops, the correlation function involves more complicated conformal integrals, the so-called `easy' and `hard' integrals, which cannot be expressed in terms of the functions \re{data1}. For the `simplest' correlation function of the half-BPS operators \re{G4}, such complicated integrals appear starting from $K+1$ loops. This explains why the relations \re{G=O2} and \re{OctagonLadders} simplify as $K\to \infty$. 
  
Recently, a nonperturbative formulation of the octagon has been proposed in Refs.~\cite{Coronado:2018ypq,Coronado:2018cxj,Kostov:2019stn,Kostov:2019auq}. It relies on a representation 
of the octagon as a product of two hexagon form factors glued together along the common side (shown by the dashed line in Fig.\ \ref{CorrelationOctagonFig}).
In this way, the octagon $\mathbb{O}(z,\bar z)$ can be expressed as an infinite sum over the mirror particles propagating between the two 
hexagons. This sum can be performed at finite value of 't Hooft coupling leading to a representation for $\mathbb{O}$ in terms of a 
 determinant of a certain semi-infinite matrix. At weak coupling, to any given order in $g^2$, the matrix becomes effectively finite-dimensional and calculation of its determinant leads to \re{OctagonLadders}. 
 
Computing the octagon for finite value of 't Hooft coupling and arbitrary kinematic variables $z$ and $\bar z$ remains an open problem.
In this paper, we exploit the determinant representation of the octagon to derive the exact expression for $\mathbb{O}(z,\bar z)$  in the 
light-like limit  $x_{12}^2,x_{13}^2,x_{24}^2,x_{34}^2\to 0$ mentioned above. We show below that the asymptotic behavior of the octagon in this limit has a remarkably simple form -- it
is governed by two functions of the coupling constant which admit a closed representation.

In terms of the cross ratios \re{cross}, the light-like limit $x_{12}^2,x_{13}^2,x_{24}^2,x_{34}^2\to 0$ corresponds to sending   $z\to 0_-$ and $\bar z\to -\infty$. \footnote{$z$ and $\bar z$ are chosen to be negative in order 
to preserve the positivity of the cross ratios \re{cross}.} It is convenient to introduce a parameterization 
\begin{align}\label{y-xi}
z=-\e^{-y-\xi} \,,\qqqquad \bar z  = -\e^{y-\xi}
\, ,
\end{align}
so that the light-like limit in question corresponds to $y \to\infty$ and $\xi$ kept fixed. Examining \re{OctagonLadders} in this limit, one finds that the octagon has a Sudakov-like form  at weak coupling \cite{Coronado:2018cxj,Kostov:2019auq}
\begin{align}\label{LC} 
\log \mathbb{O}(z,\bar z) \stackrel{y \to \infty}{=} -  {y^2\over 2 \pi^2} \Gamma(g) + g^2 \xi^2 + \frac18 C(g)  + O(\e^{-y})\,,
\end{align}
where $y=\log(\bar z/z)/2$ and $\xi=-\log(z \bar z)/2$. Here the last term on the right-hand side denotes corrections suppressed by powers of 
$z$ and $1/\bar z$. 
Notice that the dependence of $\log \mathbb{O}(z,\bar z)$ on $\xi^2$
is one-loop exact in 't Hooft coupling. Making use of this fact we can simplify our analysis
by putting $\xi=0$ from the start.

A nontrivial dependence of \re{LC} on the coupling resides in the two functions  $\Gamma(g)$ and $C(g)$. In what follows, we prove \re{LC} 
starting from the determinant representation of the octagon and derive a closed-form expression for these functions~\footnote{In the  notations of Ref.\ \cite{Coronado:2018cxj}, we have $\widetilde \Gamma(g)=g^2/4+\Gamma(g)/(8\pi^2)$.}
\begin{align}
\label{Gam} 
\Gamma(g) =   {\log (\cosh (2 \pi  g))} 
\, , \qqqquad
C(g) =- \log \left(\frac{\sinh (4 \pi  g)}{4 \pi  g}\right)
\, .
\end{align}
These relations represent the main result of the paper.
In the expression for the logarithm of the octagon \re{LC}, the function $\Gamma(g)$ is accompanied by a double-logarithmic $\log^2(\bar z/z)$ term and it is analogous to the cusp anomalous dimension. The function $C(g)$ defines a constant, $y-$independent contribution and it is known as the hard function. It is interesting to notice that it is expressible in terms of the function $\Gamma(g)$
\begin{align}
\label{ModeExpC}
C(g) = - \sum_{n = 0}^\infty \Gamma (g/2^n)
\, .
\end{align}

To appreciate the remarkable simplicity of the relations \re{LC} and \re{Gam}, we compare them with analogous equations describing the light-like asymptotics of the four-point correlation function of the simplest half-BPS operators of length $K=2$ \cite{Alday:2010zy,Alday:2013cwa,Korchemsky:2019nzm}
\begin{align}\label{cusp}
 \vev{O_2 O_2 O_2 O_2}  \sim H(g) \int_0^\infty {dy_1  \over y_1 } { dy_2\over  y_2} f(y_1)f(y_2) \e^{-S(-z/y_1,1/(-\bar z y_2))}\,,
\end{align}
where $H(g)$ is a hard function analogous to $C(g)$,  $f(y)=2 y K_0(2\sqrt{y})$ involves the modified Bessel function of the second kind
and $S(u,v)$ is the logarithm of (an appropraitely regularized) rectangular light-like Wilson loop 
\begin{align}
S(u,v) =  \frac12 \Gamma_{\rm cusp}(g) \log u \log v - \frac 12\Gamma_{\rm v}(g) \log (uv)\,.
\end{align}
The double integral in \re{cusp} comes about as a result of a resummation of the contribution of twist-two operators with large spin propagating in different OPE channels. Invoking the first-quantized picture for the correlation function \cite{Alday:2010zy}, it can also be interpreted as describing the recoil of a fast scalar particle  propagating along the null rectangle due to its interaction with the emitted radiation. The functions $\Gamma_{\rm cusp}(g)$ and $\Gamma_{\rm v}(g)$
define the large spin limit of the scaling dimensions of twist-two operators. They do not admit a closed representation in planar $\mathcal N=4$ SYM but they can be found for arbitrary 't Hooft coupling from integrability.

In the case of the `simplest' correlation function, the light-like asymptotics \re{LC} comes from the contribution of high-spin operators with twist $K$. In distinction to the $K=2$ case, the spectrum of high-twist operators is degenerate -- the number of operators with the same spin and twist grows very rapidly with $K$. As a result, for $K\to\infty$ the asymptotic behavior \re{LC} arises from the averaging over an infinite number of states propagating in different OPE channels. In the first quantized picture, the relation \re{LC} describes the propagation of $K/2$ fast particles along the null rectangle. For $K\to \infty$ these particles have an infinite energy and the recoil effect is negligible. This explains why \re{LC} has much simpler form as compared to \re{cusp}.
 
It is interesting to note that at strong coupling, i.e., for $g\to\infty$, the functions \re{Gam} grow linearly with $g$. Such behavior is exactly what one would expect from the dual semiclassical description of the 
correlation function \re{G4} in AdS/CFT. The term $g^2\xi^2$ in the right-hand side of \re{LC} seems to invalidate this behavior. We would like to emphasize that the relation \re{LC} holds up to corrections from 
operators of twist $(K+2)$ and higher. At weak coupling, these corrections are exponentially suppressed at large $y$. At strong coupling, anomalous dimensions of the contributing operators grow indefinitely 
with the coupling and their classification with respect to twist becomes redundant. It is possible to show (see Ref.~\cite{Belitsky:2020qrm}) that for sufficiently large coupling $g_\star \sim \log y/(2\pi)$, the last term in the 
right-hand side of \re{LC} becomes comparable with the first three terms. The relation \re{LC} holds for $g\ll g_\star$ whereas for $g\gg g_\star$ the octagon possesses the anticipated semiclassical AdS/CFT 
behavior~\cite{Bargheer:2019exp}.

Our subsequent presentation is organized as follows. In Section \ref{OctagonFredholmSection}, we examine the determinant representation of the octagon  in the light-like limit. We show that the problem can be reduced to computing the Fredholm determinant of an integral operator defined on a semi-infinite line with the so-called  Bessel kernel modified by a Fermi-Dirac like distribution function. This determinant is remarkably similar to those that appeared in the study of two-point correlation functions in exactly solvable models at finite temperature \cite{Its:1990,Korepin:1993kvr} and level spacing distributions in 
random matrix models \cite{Mehta,Tracy:1993xj}. In the former case, identifying
the coupling constant $g$ and its product with the kinematical variable $yg$ as the temperature and the chemical potential, respectively, we relate the weak coupling expansion of the octagon  to low temperature expansion of a certain two-point correlation function.
In Section \ref{TWSection}, we follow the approach of Refs.~\cite{Its:1990,Korepin:1993kvr,Tracy:1993xj} to derive a system of nonlinear differential equations for the Fredholm determinant under consideration. We solve these equations in Section \ref{RecursionSection} and obtain the exact result for the octagon \re{LC}  with the functions $\Gamma(g^2)$ and $C(g^2)$ given by \re{Gam}. 
Finally, we present concluding remarks in Section~\ref{ConclusionsSection} and outline directions for future work. Some details of the calculation are presented in Appendix.

%%%%%%%%%%%%%%%%%%%%%%%%%%%%%%%%%%%%%%%%%%%%%%%%%%%%%%%%%%%%%%%%%%%%%%%%%%%
\section{Octagon as a Fredholm determinant}
\label{OctagonFredholmSection}
%%%%%%%%%%%%%%%%%%%%%%%%%%%%%%%%%%%%%%%%%%%%%%%%%%%%%%%%%%%%%%%%%%%%%%%%%%%

In this section, we derive a representation of the octagon in the light-like limit as a Fredholm determinant of a certain integral operator on the semi-infinite line.

Following the hexagonalization approach \cite{Basso:2015zoa}, the octagon $\mathbb{O} (z,\bar z)$ can be expanded into an infinite sum over the virtual multi-particle states propagating on the two-dimensional worldsheet across the bridge of zero length (see Fig.~\ref{CorrelationOctagonFig}). The contribution of each state factorizes into the product of two identical hexagon form factors which are known exactly from integrability~\cite{Coronado:2018ypq,Coronado:2018cxj}. It was shown in Refs.~\cite{Kostov:2019stn,Kostov:2019auq} that, truncating the sum over the number of exchanged virtual particles to $N$, the (square of the) octagon 
can be computed as a determinant of an $N \times N$ matrix 
$(\1 - \lambda \mathbb{C} \mathbb{K})$. Here  $\lambda = -2 (\cosh y+1)$ is the overall coefficient and 
$\mathbb{C} = \{ C_{nm} \}_{0 \leq n,m \leq N}$ is a constant matrix with the only nonzero entries located on super- and subdiagonals
\begin{align}
C_{nm} = \delta_{n+1, m} - \delta_{n, m+1}
\, .
\end{align}
The matrix $\mathbb{K}  = \{ K_{nm} \}_{0 \leq n,m \leq N}$ explicitly depends on the kinematical variables $y$ and $\xi$ and the 't Hooft coupling constant, 
\begin{align}\label{K}
K_{mn} =  - {g\over 2i}\int_{|\xi|}^\infty dt {\lr{i\sqrt{t+\xi\over t-\xi}}^{m-n} - \lr{i\sqrt{t+\xi\over t-\xi}}^{n-m} \over \cosh y+\cosh t} J_m(2g\sqrt{t^2-\xi^2})J_n(2g\sqrt{t^2-\xi^2})
\, ,
\end{align}
where $J_n$ is the Bessel function.

Expanding  $\sqrt{\det (\1 - \lambda \mathbb{C} \mathbb{K})}$ in powers of the coupling constant and carrying out the integration,  one reproduces the ladder integral representation for the octagon \re{OctagonLadders} up to corrections of order $O(g^{2N})$. 
It was found  \cite{Coronado:2018ypq} to be
in complete in agreement with more traditional approaches based on the the use of (super) conformal symmetries and analytic bootstrap \cite{Chicherin:2015edu}. Refraining from the perturbative expansion and formally sending $N \to \infty$, one obtains the determinant representation for the  octagon \cite{Kostov:2019stn},
\begin{align}
\label{detOsquared}
\mathbb{O} (z,\bar z) = \lim_{N \to \infty}\sqrt{ \det (\1 - \lambda \mathbb{C} \mathbb{K}) }
\, .
\end{align}
In the following, we tacitly assume that the limit exists and drop the limit sign.

%%%%%%%%%%%%%%%%%%%%%%%%%%%%%%%%%%%%%%%%%%%%%%%%%%%%%%%%%%%%%%%%%%%%%%%%%%%
\subsection{Light-cone limit}
\label{LCSection}
%%%%%%%%%%%%%%%%%%%%%%%%%%%%%%%%%%%%%%%%%%%%%%%%%%%%%%%%%%%%%%%%%%%%%%%%%%%

Going to the light-like limit, we substitute \re{y-xi} into \re{detOsquared} and send $y\to\infty$. We expect that  in this limit the relation \re{detOsquared} should reproduce the asymptotic behavior of the 
octagon \re{LC}. Aimed to find the exact expressions for the functions $\Gamma(g)$ and $C(g)$ in \re{LC}, we can simplify the analysis by setting $\xi=0$.

A close examination of the matrix elements  $\lambda K_{mn}$ shows that for $y \to \infty$ and $\xi=0$ the leading 
contribution to the integral  \re{K} stems from $t \sim y$. This allows us to approximate $\cosh t \sim {\rm e}^{t}/2$, yielding
\begin{align}
\lambda K_{mn} =  2g \sin\left(\frac{\pi}2(m-n) \right) \int_0^\infty dt {J_m(2gt)J_n(2gt)\over 1 +  \e^{t-y}}
\, .
\end{align}
Then, the matrix $k_{nm}= \lambda (CK)_{nm} = \lambda (K_{n+1,m}-K_{n-1,m})$ has nonzero entries only if $n$ and $m$ have the same parity. This suggests to split the matrix into two irreducible blocks $(k_{+})_{nm}=k_{2n,2m}$ and $(k_-)_{nm}=k_{2n+1,2m+1}$ with
\begin{align}\notag
\label{hh}
& (k_+)_{nm} = (-1)^{n+m} 4n \int_0^\infty {dt\over t} { J_{2n}(2gt) J_{2m}(2gt)\over \e^{t-y}+1} +\delta_{n,0} (-1)^m 2g \int_0^\infty {dt } 
{J_{1}(2gt) J_{2m}(2gt)\over \e^{t-y}+1}\,,
\\
& (k_-)_{nm} = (-1)^{n+m} 2(2n+1) \int_0^\infty {dt\over t} {J_{2n+1}(2gt) J_{2m+1}(2gt)\over \e^{t-y}+1} \,.
\end{align}
The two infinite-dimensional matrices $\Bbbk_\pm  = \{ (k_\pm)_{nm} \}_{0 \leq n,m < \infty}$
 are in fact related to each other through a similarity transformation\footnote{This can be shown using the property of Bessel functions $2(n+1) J_{n+1}(z) = z\lr{J_n(z) + J_{n+2}(z)}$.}
\begin{align}
\mathbb{U}^{-1} \, \Bbbk_- \, \mathbb{U} =  \Bbbk_+
\, ,
\end{align}
where $\mathbb{U}$ is a lower-triangular matrix with $U_{nm}=(2n+1)$ for $n\ge m\ge 0$. Then, we use the relation
$\det (\1 -\lambda \mathbb{C} \mathbb{K}) =\det (\1 - \Bbbk_+)\det (\1 - \Bbbk_-)=[\det (\1 - \Bbbk_-)]^2$ to find the following representation for the octagon \re{detOsquared} in the light-like limit
\begin{align}
\label{OctagonDetkMinus}
\mathbb{O} = \det (\1 - \Bbbk_-)
\, .
\end{align}
As we show in a moment, the expression on the right-hand side of \re{OctagonDetkMinus} can be converted into the Fredholm determinant of an integral operator.

%%%%%%%%%%%%%%%%%%%%%%%%%%%%%%%%%%%%%%%%%%%%%%%%%%%%%%%%%%%%%%%%%%%%%%%%%%%
\subsection{Weak coupling}
\label{WeakCouplingSection}
%%%%%%%%%%%%%%%%%%%%%%%%%%%%%%%%%%%%%%%%%%%%%%%%%%%%%%%%%%%%%%%%%%%%%%%%%%%

Let us first demonstrate how the relation \re{OctagonDetkMinus} works at weak coupling. We apply the identity
\begin{align}
\label{G-tr}
\log \mathbb{O} = \tr \log (\1- \Bbbk_-) = - \tr(\Bbbk_-) - \ft12  \tr(\Bbbk_-^2) - \ft13  \tr(\Bbbk_-^3) + \dots
\end{align}
and take into account that the matrix elements \re{hh} scale at weak coupling as $(k_-)_{nm}\sim (g^{2})^{n+m+1}$ in virtue of $J_n(2gt) \sim (2gt)^n$. As a consequence,  the leading contribution to \re{G-tr} comes from the matrix element
$(k_-)_{00}$  
\begin{align}\notag\label{G-1loop}
\log  \mathbb{O} =  - (k_-)_{00} + O(g^4)  &= - 2 g^2 \int_0^\infty  \frac{ dt\, t}{e^{t-y}+1} + O(g^4)  
\\
 &= -\lr{ y^2+{\pi^2\over 3} } g^2 + O(g^4) + O(\e^{-y})
\, .
\end{align}
This relation has the expected form \re{LC} and it can be used to identify lowest order form of the functions $\Gamma(g)$ and $C(g)$. 

It is straightforward to compute higher order corrections to \re{G-tr}. Expanding the matrix elements \re{hh} in powers of the coupling constant, we find that at large $y$ they scale as $(k_-)_{nm}\sim g^{2L} (y^{2L} + O(y^{2L-2}))$ with $L=n+m+1$. This relation implies that
each term on the right-hand side of \re{G-tr} has similar asymptotic behavior, i.e. the power of $y^2$ increases with the order in $g^2$. Examining the sum of all terms we find however that the corrections to \re{G-tr}  scales as $y^2$ to any order in the coupling. This comes about as a result of massive cancellations between  $y^4, y^6, \dots$ contributions from various terms in  \re{G-tr}. A question arises what a priori features of \re{OctagonDetkMinus} stop   $\log  \mathbb{O}$  from receiving higher order effects 
in $y^2$? We shall answer it in the subsequent analysis by mapping this property into an analogous property of correlation functions
in log-gases and random matrices~\cite{Forrester}.

Matching the resulting weak coupling expansion of \re{G-tr} to the expected form \re{LC} we obtain perturbative expressions for the
functions $\Gamma (g)$ and $C (g)$,
 \begin{align}\notag\label{Gamma-weak}
&\Gamma(g) =2 \pi ^2 g^2-\frac{4 \pi ^4 g^4}{3}+\frac{64 \pi ^6 g^6}{45}-\frac{544 \pi ^8 g^8}{315}
+
O\left(g^{10}\right)\,,
\\[2mm]
& C(g) =-\frac{8 \pi ^2 g^2}{3}+\frac{64 \pi ^4 g^4}{45}-\frac{4096 \pi ^6 g^6}{2835}+\frac{8192 \pi ^8 g^8}{4725}+O\left(g^{10}\right)
\, .
\end{align}
These relations are in agreement with the results of Refs.\ \cite{Coronado:2018ypq,Coronado:2018cxj,Kostov:2019stn}.  

%%%%%%%%%%%%%%%%%%%%%%%%%%%%%%%%%%%%%%%%%%%%%%%%%%%%%%%%%%%%%%%%%%%%%%%%%%%
\subsection{Relation to Bessel kernel}
\label{BesselKernelSection}
%%%%%%%%%%%%%%%%%%%%%%%%%%%%%%%%%%%%%%%%%%%%%%%%%%%%%%%%%%%%%%%%%%%%%%%%%%%

The relation \re{OctagonDetkMinus} yields the null octagon in the form of a determinant of a semi-infinite matrix. 
It proves instructive to rewrite it as a Fredholm 
determinant of an integral operator represented by the matrix $\Bbbk_-$ defined in \re{hh}. To accomplish this goal, we first examine 
the square of this matrix
\begin{align} 
\label{tr-h2}
(\Bbbk_-^2)_{nk} 
= 2(2n+1)(-1)^{n+k} \int_0^\infty {dt_1 \over t_1}  {J_{2n+1}(2gt_1) \over \e^{t_1-y}+1}
\int_0^\infty {dt_2 \over t_2} {J_{2k+1}(2gt_2)\over \e^{t_2-y}+1} H(2gt_1,2gt_2)\,.
\end{align} 
Here we replaced $\Bbbk_-$ with its explicit expression \re{hh} and introduced a notation for the function
\begin{align}\notag
\label{H-def}
H(2gt_1,2gt_2)
& =\sum_{m\ge 0} 2(2m+1)  J_{2m+1}(2gt_1)J_{2m+1}(2gt_2) 
\\
&=
\frac{2 g t_1 t_2 \left(t_1 J_0\left(2 g t_2\right) J_1\left(2 g t_1\right)-t_2  J_0\left(2 g t_1\right) J_1\left(2 g t_2\right)\right)}{t_1^2-t_2^2}
\, .  
\end{align}
It approaches a finite value at the coinciding points and is odd in $t_1$ and $t_2$ separately.
Using the properties of the Bessel functions it is straightforward to verify that it satisfies the following relations
\begin{align}\notag
& \int_0^\infty {dt_2\over t_2} H(2gt_1,2gt_2) H(2gt_2,2gt_3) = H(2gt_1,2gt_3)
\, , \\
& \int_0^\infty {dt_2\over t_2} H(2gt_1,2gt_2) J_{2n+1}(2gt_2) = J_{2n+1}(2gt_1)
\, .
\end{align}  

Repeatedly applying \re{tr-h2} we can evaluate $\tr( \Bbbk_-^L)$ as a $L-$folded integral involving the function \re{H-def} 
\begin{align}
\tr( \Bbbk_-^L) = \int_0^\infty {dt_1 \over t_1} \dots \int_0^\infty {dt_L \over t_L} {H(2gt_1,2gt_2)\dots H(2gt_L,2gt_1)\over (\e^{t_1-y}+1)\dots (\e^{t_L-y}+1)}
\, .
\end{align}
Substituting this relation into \re{G-tr}, we can express the null octagon as a Fredholm determinant 
\begin{align}\label{G-detH}
\mathbb{O} = \det (\1-\mathbb{H}) \,,
\end{align}
where the integral operator $\mathbb{H}$ is defined on the semi-infinite line through its action on  an arbitrary test function $f(t)$
\begin{align}\label{calH}
\mathbb{H} f(t) = \int_0^\infty {dt' \over t'} {H(2gt,2gt')\over \e^{t'-y}+1}f(t')
\, .
\end{align}    
At large $y$, the exponential factor in the denominator suppresses the contribution from $t'\gg y$ and serves as an ultraviolet cut-off.

The relation \re{calH} is remarkably close to the definition of the so-called Bessel kernel that appeared in the study of  
level spacing distributions in
random matrices, see, e.g., the volume \cite{Harnard2011} 
for an overview. This kernel is defined as
\begin{align}\label{Bessel}
K_\beta (x_1,x_2) = {\phi_\beta (x_1)\psi_\beta (x_2)-\psi_\beta (x_1)\phi_\beta (x_2)\over x_1-x_2} = \frac14\int_0^1 dt\, \phi_\beta (x_1 t)\phi_\beta (x_2 t)\,,
\end{align}
where $\phi_\beta (x) = J_\beta (\sqrt{x})$ and $\psi_\beta (x) = x\phi_\beta'(x)$ are expressed in terms of the Bessel function of order $\beta$. For $\beta=0$ it admits the form
\begin{align}\label{K0}
K(x_1,x_2) =\frac{\sqrt{x_1} J_1\left(\sqrt{x_1}\right) J_0\left(\sqrt{x_2}\right)-\sqrt{x_2} J_0\left(\sqrt{x_1}\right) J_1\left(\sqrt{x_2}\right)}{2 (x_1-x_2)}
\, ,
\end{align}
with obvious decluttering notation for $K(x_1,x_2) = K_{\beta = 0}(x_1,x_2)$. 
Comparing the relations \re{H-def} and \re{K0}, we notice that the two functions coincide upon appropriate identification of the variables
\begin{align}
H (2gt_1,2gt_2) = 2 \sqrt{x_1 x_2} \,K(x_1,x_2) \,, 
\end{align}
where $x_i = (2g t_i)^2$. 

To accommodate for the additional factor in the denominator of \re{calH}, we modify the Bessel kernel $K(x_1,x_2)$ by multiplying it by the 
cut-off function $\chi(x_2)$ defined as
\begin{align}\label{chi}
\chi(x) =  {1\over \e^{(\sqrt{x}-\sqrt{s})/(2g)}+1} \,, 
\end{align} 
where we introduced the variable $s= 4 g^2 y^2$. The resulting integral operator looks as
\begin{align}\label{calK}
\mathbb{K}_\chi f(x) = \int_0^\infty {dx'} \, K(x,x')\chi(x') f(x')
\, ,
\end{align}
with $f(x)$ being a test function. 
Upon identification of the coordinates $x=(2g t)^2$, the operators \re{calH} and \re{calK} are related to each other by a similarity transformation $\mathbb{H}= \mathbbm{x}^{1/2} \,\mathbb{K}_\chi \,  \mathbbm{x}^{-1/2}$ involving 
the coordinate operator $\mathbbm{x}$. Their Fredholm determinants coincide and we conclude that  
\begin{align}\label{G-detK}
\mathbb{O} = \det (\1-\mathbb{K}_\chi) 
\, .
\end{align}
In this representation, the dependence of the null octagon on the coupling constant $g$ and the cross ratio $y$ resides in the function $\chi$ entering the definition \re{calK} 
of the integral operator $\mathbb{K}_\chi$.    

The function \re{chi} looks very similar to the Fermi-Dirac distribution. Slightly abusing the definition, we can interpret the 't Hooft coupling $g$ as a temperature and the variable $\sqrt{s}=2g y$ 
as a chemical potential. The function $\chi(x)$ approaches $1$ for $x\ll s$ and vanishes for 
$x\gg s$. At zero temperature, or equivalently in the weak coupling limit,  $g^2\to 0$, and fixed chemical potential $s= 4 g^2 y^2$, it becomes a step function 
\begin{align}\label{step}
\lim_{g^2\to 0,\atop  \text{$s$ fixed}} \chi(x) = \theta(s-x)\,.
\end{align}
In this case, the operator \re{calK} acts on a finite interval $[0,s]$ and its  
Fredholm determinant  takes a remarkably simple form \cite{Tracy:1993xj}
\begin{align}
\label{G-detK0}
\lim_{g^2\to 0,\atop  \text{$s$ fixed}}  \log \det (\1-\mathbb{K}_\chi) = -s/4 \,.
\end{align}
Together with \re{G-detK} this provides a prediction for the null octagon in the limit $g^2\to 0$ with $s=4g^2 y^2$ kept fixed. Indeed,  we find in this limit from \re{LC}  that 
$\log \mathbb{O} = - g^2 y^2 $ in perfect agreement with \re{G-detK} and \re{G-detK0}.

For finite values of the 't Hooft coupling, the function $\chi(x)$ is different from the step function \re{step}. The integral operator \re{calK}  acts on the semi-infinite 
axis and its Fredholm determinant is different from \re{G-detK0}. To best of our knowledge, 
it has not been discussed in the literature before.

Comparing the relations
\re{G-detK} and \re{LC}, we expect that the logarithm of the Fredholm determinant of the operator \re{calK} should take the following form
\begin{align}
\label{goal} 
\log \det (\1-\mathbb{K}_\chi)  = - \frac{s}{8\pi^2 g^2} \Gamma(g) + \frac18 C(g)  + O(\e^{-\sqrt{s}/(2g)})\,,
\end{align}    
where $s=4g^2 y^2 \gg 1$ and  $g$ is kept fixed.
Our goal is to prove this relation and to calculate the functions of the coupling constant which enter its right-hand side. Comparing \re{goal} to \re{G-detK0}, we observe that the only 
modification from the change of the step function \re{step} to the Fermi-Dirac distribution \re{chi} is that the latter merely induces two nontrivial functions of the 't Hooft constant but does not modify 
the linear dependence on the variable $s$.

%%%%%%%%%%%%%%%%%%%%%%%%%%%%%%%%%%%%%%%%%%%%%%%%%%%%%%%%%%%%%%%%%%%%%%%%%%%
\section{Method of differential equations}
\label{TWSection}
%%%%%%%%%%%%%%%%%%%%%%%%%%%%%%%%%%%%%%%%%%%%%%%%%%%%%%%%%%%%%%%%%%%%%%%%%%%

As was already mentioned, Fredholm determinants naturally arise in the study of two-point correlation functions in exactly solvable models \cite{Korepin:1993kvr}.
A general method for evaluating such determinants has been developed in Ref.~\cite{Its:1990}. It reduces the problem to solving a system of nonlinear differential equations. The Fredholm determinant plays the role of a $\tau-$function for this system of equations. 
At zero temperature, it can be expressed in terms of Painlev\'e transcendents, whereas at nonzero temperature, it becomes
possible to find its asymptotic behavior for different values of the parameters. 
In the special case of the Bessel kernel, this method has been applied in Ref.~\cite{Tracy:1993xj} to compute the Fredholm determinant at zero temperature.

In this section, we apply the approach of Refs.\ \cite{Its:1990,Korepin:1993kvr,Tracy:1993xj} to evaluating the Fredholm determinant of the operator \re{calK}. Despite the fact that the operator \re{calK} is different from those studied in Refs.\ \cite{Its:1990,Korepin:1993kvr,Tracy:1993xj}, we are able to derive a corresponding system of differential equations and, then, solve it exactly.

It proves convenient to pull out the dependence of the operator $\mathbb{K}_\chi$ in  \re{calK} on the cut-off function $\chi(x)$. By introducing an operator $\bbchi$ with a diagonal kernel $\chi(x_1)\delta(x_1-x_2)$, the operator \re{calK} can be rewritten in the factorized form $ \mathbb{K}_\chi = \mathbb{K}  \bbchi$. Here $\mathbb{K}$ is given by \re{calK} with the function $\chi(x)$ 
replaced by $1$. Introducing the notation for the logarithm of the octagon
\begin{align}\label{d} 
d(y,g) \equiv  \log  \det(\1- \mathbb{K}  \bbchi) 
\,,
\end{align}
we observe that its dependence on $y$ and $g$ now resides in $\bbchi$, or equivalently in the function $\chi(x)$ defined in \re{chi}. It is easy to see that this function satisfies the relation
\begin{align}\label{chi-rel}
x\partial_x \chi   =-\frac12 g\partial_g \chi \,,
\end{align}
which plays an important role in what follows.

\subsection{Auxiliary functions and potentials}

To obtain differential equations for $d(y,g)$, we first compute its derivatives with respect to $y=\sqrt{s}/(2g)$ and $g$.
Since the operator $\mathbb{K}$ depends on neither of these variables, the derivatives solely land on $\bbchi$.  
In order to avoid proliferation of formulas, we use a common notation $\alpha=\{y,g\}$ for the two parameters. Then, the derivative of \re{d} with respect 
to $\alpha$ is given by
\begin{align}\label{der}
\partial_\alpha d(s,g) = -\int _0^\infty dx\, R(x,x) \partial_\alpha  \chi(x)  \,,
\end{align}
where $R(x,x)= \lim_{y\to x} R(x,y)$ is defined as a limiting value of the kernel of the operator~\footnote{We use here quantum mechanical notations for the eigenstates $\ket{x}$ of the coordinate operator $\mathbbm{x}$ on the 
semi-infinite line $0\le x < \infty$.}
\begin{align}\label{R-def}
R(x,y) = \vev{x|{1\over 1- \mathbb{K} \bbchi} \mathbb{K}   | y}\,.
\end{align}
Having determined $R(x,x)$, we can apply \re{der} to compute the Fredholm determinant.

Following Refs.\ \cite{Its:1990,Korepin:1993kvr,Tracy:1993xj}, we introduce two auxiliary functions
\begin{align} \label{PQ} 
Q(x) = \vev{x|{1\over \1- \mathbb{K} \bbchi}  | \phi}\,, 
\qqqquad 
P(x) = \vev{x|{1\over \1- \mathbb{K} \bbchi} | \psi}\,, 
\end{align}
where $\phi (x) \equiv \vev{x|\phi} = J_0(\sqrt{x})$ and $\psi (x) \equiv \vev{x|\psi} = -\frac12 \sqrt{x} J_1(\sqrt{x})$ are the functions defining the Bessel kernel \re{Bessel} for $\beta=0$. The rationale for 
introducing \re{PQ}  is that the kernel $R(x,y)$ introduced in \re{R-def}  admits a concise representation in terms of these functions
\begin{align}
R(x,y) = {Q(x) P(y) - P(x) Q(y)\over x-y}\,.
\end{align}
In the limit $x\to y$, we then obtain
\begin{align}\label{Rxx}
R(x,x) = Q(x) P'(x) -P(x) Q'(x)\,,
\end{align}
where the prime stands for the derivative with respect to $x$.

Our goal now is to derive differential equations for the functions $Q(x)$ and $P(x)$ introduced in \re{PQ}.
In the limiting case of the step function \re{step}, this has been done in Ref.\  \cite{Tracy:1993xj}. We show below that the method can 
be extended to the general case of the function $\chi(x)$ given by \re{chi}.

Using the definition \re{PQ} and taking into account the relation \re{chi-rel}, it is possible to show (see Appendix for details) that 
the functions $Q(x)$ and $P(x)$ obey a system of first order linear differential equations 
\begin{align}\notag\label{der2}
& x \partial_x Q(x)= P(x) + \frac14 u Q(x) -\frac12 g\partial_g Q(x)\,,
\\
& x \partial_x P(x)=-{x\over 4} Q(x) + \frac12 v Q(x)-\frac14 u P(x) -\frac12 g\partial_g P(x)\,,
\end{align}
where it is tacitly assumed that  $Q(x)$ and $P(x)$ also depend on $g$ and $y$. Here $u$ and $v$ are the so-called potentials in terminology of Refs.~\cite{Its:1990,Korepin:1993kvr}. They are given by the following matrix elements
\begin{align}\label{u-def}
u = \vev{\phi|  \bbchi {1\over \1- \mathbb{K} \bbchi} |\phi}\,, \qqqquad
 v = \vev{\phi|  \bbchi {1\over \1- \mathbb{K} \bbchi} |\psi}\,.
\end{align}
The two sets of the variables, $Q,P$ and $u,v$ are not independent of each other. Differentiating $u$ and $v$ with respect to $\alpha=\{y,g\}$, we deduce  
\begin{align}\label{der1}\notag
& \partial_\alpha u =  \int_0^\infty dx \, Q^2(x)  \partial_\alpha\chi (x)\,,
%\qqqquad
\\
& \partial_\alpha v =   \int_0^\infty dx \, Q(x) P(x)  \partial_\alpha\chi (x)\,,
\end{align}
Being combined together, the relations \re{der2} and \re{der1} define a closed system of integro-differential equations.
 
Applying the equations \re{der2} and \re{der1}, we find that the potentials satisfy the following relation (see Appendix)
\begin{align}\label{uv-rel}
 v+\frac18 u^2+\frac12 u - \frac14g \partial_g  u =0\,.
\end{align}
Excluding $P(x)$ from \re{der2} and taking into account \re{uv-rel}, we derive a second-order differential equation for the function $Q(x)$ 
\begin{align}\label{pde}
\left(g\partial_g + 2 x\partial_x \right)^2 Q(x)+ \left(x-g \partial_g u+u\right)Q(x) = 0\,.
\end{align}
In a similar manner, we can exclude $P(x)$ from \re{Rxx} to get
\begin{align}\label{R-Q}\notag
R(x,x) 
&= \frac12
\big[  \partial_x Q(x) (g \partial_g  +2 x\partial_x) Q(x)  
 -  Q(x)  \partial_x (g \partial_g  +2 x\partial_x) Q(x)  \big]
\\
& = - \frac12 Q^2(x)  \partial_x (g \partial_g  +2 x\partial_x) \log Q(x) 
 \,.
\end{align}
Solving \re{pde} for the function $Q(x)$ and substituting the solution to \re{R-Q} and \re{der}, we can
express the derivative of the Fredholm determinant on the left-hand side of \re{der} in terms of a single function $u=u(y,g)$, but  it still needs to be determined.

\subsection{Fredholm determinant from the potential}

An additional condition for $u$ arises from the relation
\begin{align}
\label{RtoQQ}
\frac14 Q^2(x)  = \lr{1+x \partial_x  +\ft12 g\partial_g}  R(x,x)  
\, ,
\end{align}
whose derivation can be found in the Appendix. Multiplying both sides of \re{RtoQQ} by $\partial_\alpha  \chi (x)$ and integrating them with 
respect to $x$ over the positive half line, we get  
\begin{align}\notag
\frac14 \partial_\alpha u &=\int_0^\infty dx \left[\partial_x \lr{x R(x,x) }+\frac12 g\partial_g  R(x,x)   \right]\partial_\alpha  \chi (x) 
\\
&= \frac12 g\partial_g \int_0^\infty dx R(x,x)   \partial_\alpha   \chi(x) +  \frac12   \int_0^\infty dx R(x,x)  [ \partial_\alpha,g\partial_g ]   \chi (x)  \,.
\end{align}
Here in the first relation, we used \re{der1} to replace the integral on the left-hand side with $\partial_\alpha u /4$. The second relation was obtained by integrating by parts the first term inside the brackets in the first line and subsequently 
applying \re{chi-rel}. Then, we can use \re{der} to re-express the integrals in the second line of the last relation in terms of derivatives of the Fredholm determinant $\partial_\alpha d$. In this way, we get for $\alpha=\{y,g\}$
\begin{align}\label{eqs}
\partial_y \left[  g\partial_g   d(y,g) + \frac12 u\right]= \partial_g\left[  g\partial_g   d(y,g) + \frac12 u\right]=0\,.
\end{align}
Solving these equations we find
\begin{align}\label{u-eq}
u=-2g\partial_g   d(y,g)
\, .
\end{align}
Obviously, a general solution to \re{eqs} contains an arbitrary integration constant. We can show that it equals zero by going to the limit
$g\to 0$ with $y$ kept large but fixed. In this limit, the function $\chi(x)$ reduces to the step function \re{step} with $s=4g^2y^2\to 0$.
Because the interval $[0,s]$ shrinks into a point, the expressions on both sides of the last relation vanish simultaneously leaving no room for the integration constant.
 
 Finally, integrating \re{u-eq} we obtain the expression for the Fredholm determinant in 
terms of the potential $u(y,g)$ 
\begin{align}\label{D-u}
d(y,g)=- \frac12 \int_0^g {d g'\over g'} u(y,g') \,.
\end{align}
As before the integration constant is fixed from the requirement for both sides to vanish as $g\to 0$. 

Equation \re{u-eq} also leads to a nontrivial relation between the potential $u$ and the function $Q(x)$. Applying \re{der} for $\alpha=g$ and replacing $R(x,x)$ with its expression \re{R-Q}, we find from \re{u-eq}
\begin{align}\label{u-int}
u
= - \int_0^\infty dx\, Q^2(x)   \partial_x (g \partial_g  +2 x\partial_x) \log Q(x)  g\partial_g  \chi(x) \,.
\end{align}
Assembling Eqs.\ \re{pde} and \re{u-int} together allows us to fix the functions $u$ and $Q(x)$. Having found $u$ we can apply \re{D-u}
and compute the Fredholm determinant.

Several comments are in order now.

We would like to emphasize that the analysis in this section does not rely on the particular form of the cut-off function \re{chi} and it only makes use of the relation \re{chi-rel}. As a consequence, the relations \re{pde} and \re{u-int}  hold for any function $\chi(x)$ depending on $\sqrt{x}/(2g)$ and $y$.

Significant simplification occurs in the zero temperature limit, $g\to 0$ with $s=4g^2 y^2$ kept fixed. In this limit, we expect to recover the well-known result for the Fredholm determinant of the Bessel kernel \cite{Tracy:1993xj}.
Using \re{chi-rel} and \re{step}, we find that $g\partial_g  \chi(x)=-2x\partial_x \chi(x) \to 2 s\delta(x-s)$ and the
integral on the right-hand side of \re{u-int} is localized at $x=s$. Then, the relations \re{u-int} and \re{pde} yield a differential equation for $u=u(s)$ that can be identified 
as of Painlev\'e III type, in agreement with \cite{Tracy:1993xj}.

For finite $g$, we have to deal with the integral in \re{u-int}. This complicates the analysis as compared to the zero temperature limit. 
Notice that the integrand in \re{u-int} involves a derivative of the function \re{chi}. It is easy to see from 
\re{chi} that the function  
$g\partial_g  \chi(x)$ is peaked at $s=(2gy)^2$ and decays exponentially fast for $|\sqrt{x}-\sqrt{s}|\gg 2g$. This suggests that the integral in \re{u-int} receives a 
dominant contribution from $x$ in the vicinity of $x=s$. In the next section, we use this fact to simplify the relations \re{pde} and \re{u-int}.

%%%%%%%%%%%%%%%%%%%%%%%%%%%%%%%%%%%%%%%%%%%%%%%%%%%%%%%%%%%%%%%%%%%%%%%%%%%
\section{Scaling limit and recurrence relation}
\label{RecursionSection}
%%%%%%%%%%%%%%%%%%%%%%%%%%%%%%%%%%%%%%%%%%%%%%%%%%%%%%%%%%%%%%%%%%%%%%%%%%%

To identify the leading contribution to the integral in \re{u-int}, we zoom into the region $|\sqrt{x}-\sqrt{s}|= O(2g)$ by 
changing the integration variable to $\sqrt{x}-\sqrt{s} = 2g z$, or equivalently
$z=\sqrt{x}/(2g)-y$. In order to evaluate the integral \re{u-int}, it is sufficient to know the function $Q(x)$ in this region.
We denote it as
\begin{align}\label{change}
q(z) = Q(x)
\,,\qqqquad
x=4g^2(y+z)^2 \,,
\end{align}
where the dependence on $g$ is tacitly implied. 

Taking into account the identity
\begin{align} 
& \lr{ g \partial_g  +2 x\partial_x } Q(x) = g \partial_g q(z)
\, ,
\end{align}
we obtain from \re{pde} and \re{u-int} that the function $q(z)$ satisfies the system of equations
\begin{align}
\label{DiffEqForq}
&  \left(g\partial_g  \right)^2 q(z)+ \left[4g^2(y+z)^2 -g \partial_g u +u \right]q(z)  = 0 \,,&
\\[2mm]
\label{sch}
& \int_{-\infty}^\infty dz  \frac{\e^{z}}{ ({\rm e}^{z}+1)^2}  (z+y) q^2(z)  \partial_z   \lr{g \partial_g} \log q(z) =-u\,. &
\end{align} 
Here in the second relation, we extended the lower integration limit from $z=-y$ to $z=-\infty$. This is justified at large $y$ up to exponentially small corrections $O(\e^{-y})$.

Applying the logarithmic derivative with respect to the coupling to both sides of \re{sch} and taking into account \re{DiffEqForq}, we get
\begin{align}\label{mom2}
g\partial_g u =8 g^2  \int_{-\infty}^\infty dz  \frac{ {\rm e}^{z}}{ ({\rm e}^{z}+1)^2}    (z+y)^2 q^2(z)\,.
\end{align}
We can obtain a similar relation by the change of variables \re{change} in the first relation in \re{der1}  
\begin{align}\label{mom1}
\partial_y u = 8 g^2  \int_{-\infty}^\infty dz  \frac{ {\rm e}^{z}}{ ({\rm e}^{z}+1)^2}    (z+y)  q^2(z)\,.
\end{align}
Again, it is valid up to exponentially suppressed contributions $O ({\rm e}^{-y})$. 

Comparing the expressions on the right-hand side of the last two relations, we notice that they only differ 
in the power of the variable $(z+y)$. This suggests to introduce the moments
\begin{align}\label{mom-def}
Q_\ell = \int_{-\infty}^\infty  dz  \frac{{\rm e}^{z}}{ ({\rm e}^{z}+1)^2} (z+y)^\ell  q^2(z)\,.
\end{align} 
For $\ell=1$ and $\ell=2$ we have from \re{mom2} and \re{mom1}
\begin{align}\label{Q1}
Q_1 = {1\over 8g^2} \partial_y u \,,\qqqquad Q_2={1\over 8g} \partial_g u\,.
\end{align}
The differential equation \re{sch} leads to a recurrence relation for $Q_\ell$. 
To show this, we  use \re{DiffEqForq} to derive a differential equation for the function $q^2(z)$
\begin{align}\label{diff-eq}
\left[(g\partial_g)^3+ 4 (u - g\partial_g u)  (g\partial_g) - 2 g^2\partial_g^2u \right]  q^2(z) = - 16g^2 (g\partial_g+1) (z+y)^2 q^2(z)
\, .
\end{align}
Integrating both sides with the same measure as in \re{mom-def}, we get
\begin{align}\label{mom-eq}
\left[(g\partial_g)^3+ 4 (u - g\partial_g u)  (g\partial_g) - 2 g^2\partial_g^2u \right]  Q_\ell = - 16g^2 \partial_g \lr{g \,Q_{\ell+2}}
\, .
\end{align}
Together with \re{Q1}, this relation can be used to obtain the moments $Q_\ell$ for any integer $\ell$. In particular for $\ell=0$, we find
\begin{align}\label{eq-Q0}
\left[(g\partial_g)^3+ 4 (u - g\partial_g u)  (g\partial_g) - 2 g^2\partial_g^2u \right]  Q_0 = -2 g^2\partial_g^2u \,.
\end{align}
It is easy to see that $Q_0=1$ is a special solution to \re{eq-Q0}. We show in the next subsection that it is this solution that leads to the 
expected expression for the null octagon \re{LC}.

%%%%%%%%%%%%%%%%%%%%%%%%%%%%%%%%%%%%%%%%%%%%%%%%%%%%%%%%%%%%%%%%%%%%%%%%%%%
\subsection{Weak coupling analysis}
\label{WeakSection}
%%%%%%%%%%%%%%%%%%%%%%%%%%%%%%%%%%%%%%%%%%%%%%%%%%%%%%%%%%%%%%%%%%%%%%%%%%%

Let us first examine \re{mom-eq} at weak coupling, for $g\to 0$ with $s=4g^2 y^2$ being fixed. As was explained earlier, in this (zero temperature) limit, the cut-off function $\chi$ reduces 
to the step function \re{step} and we encounter the Fredholm determinant of the Bessel kernel studied in  \cite{Tracy:1993xj}. Then, by solving  \re{DiffEqForq} and \re{sch} we expect to recover expressions obtained in \cite{Tracy:1993xj}. In our notations, they read
\begin{align}\label{triv}
q^{(0)} (z) = 1\,,\qqqquad u^{(0)}  = 4g^2 y^2\,.
\end{align} 
Here we added a superscript to all functions to indicate that these relations are valid to the leading order in 't Hooft coupling. Indeed, substituting \re{triv} into \re{mom2} and \re{mom1}, 
we verify that both relations are automatically satisfied.

Substituting the first relation in \re{triv} into \re{mom-def}, we obtain the leading correction to the moments at weak coupling
\begin{align}\label{Q=0}
Q^{(0)}_\ell =  \int_{-\infty}^\infty  dz  \frac{ {\rm e}^{z}}{ ({\rm e}^{z}+1)^2}  (z+y)^\ell\,.
\end{align}
For nonnegative integer $\ell$, the function $Q^{(0)}_\ell$ is a polynomial of degree $\ell$ in $y$.  Since the integration measure 
in \re{Q=0} is even in $z$,  this polynomial 
has a definite parity in $y$ and involves odd (even) powers of $y$ for odd (even) $\ell$. The integral on the right-hand side of \re{Q=0} can be easily computed using the identity
\begin{align}\label{Ber}
 \sum_{\ell\ge 0} {(ik)^\ell\over \ell!} Q^{(0)}_\ell =  \int_{-\infty}^\infty  dz  \frac{ {\rm e}^{z}}{ ({\rm e}^{z}+1)^2} \e^{ik(z+y)} = {k\pi \over \sinh(k\pi)}\e^{i k y}  \,.
\end{align}
Comparing the coefficients in front of the powers of $k$ on both sides, we obtain 
\begin{align}\label{Q-Ber}
Q^{(0)}_\ell =B_\ell\left(\frac12 +\frac{iy}{2\pi}\right) (-2i\pi)^\ell\,,
\end{align}
where $B_\ell(x)$ are Bernoulli polynomials. For the first few  values of $\ell$, we get from \re{Q-Ber}
\begin{align}\label{cases}
Q^{(0)}_0=1\,,\qqquad Q^{(0)}_1=y\,,\qqquad Q^{(0)}_2=y^2 + {\pi^2\over 3}\,,\qqquad Q^{(0)}_3=y( y^2 +  \pi^2) 
\, .
\end{align}
It is easy to verify using the second relation in \re{triv} that the expressions for $Q^{(0)}_1$ and $Q^{(0)}_2$ are in agreement with \re{Q1}. 

For nonzero coupling, we expect that the moments \re{cases} receive $O(g^2)$ corrections. It turns out that the lowest moment  is protected,  $Q_0=1$. The same holds true for 
all odd moments evaluated at $y=\pm i\pi$. 
\begin{align}\label{Q-zero}
Q_{2k+1} \Big|_{y=\pm i\pi }=0 \,,\qqqquad (\text{for $k=1,2,\dots$})
\, .
\end{align}
This relation implies that for arbitrary $y$ and $g$, the odd moments $Q_{2k+1}$ can be factorized into a product of $y^2+\pi^2$ and a polynomial in $y$ of degree $2k-1$.

To show the absence of quantum corrections to $Q_0$, we replace $Q_0 = 1+ g^2 Q_0^{(1)} + \dots$ in \re{eq-Q0}
and compare the coefficients in the power expansion in $g^2$ on both sides of \re{eq-Q0}. Taking into account that $u=O(g^2)$ (see \re{triv}), we find that $Q_0^{(1)} =Q_0^{(2)} =\dots =0$. 
In this way, we arrive at the relation
\begin{align}\label{Q0-norm}
Q_0 = \int_{-\infty}^\infty  dz  \frac{ {\rm e}^{z}}{ ({\rm e}^{z}+1)^2}  q^2(z)= 1\,.
\end{align}
Notice that the differential equation \re{diff-eq} is invariant under rescaling $q^2(z) \to \lambda\, q^2(z)$. The relation \re{Q0-norm} breaks this symmetry and fixes the normalization 
of $q^2(z)$.

Let us now turn to the odd moments. According to \re{Q-Ber}, for $y= i \pi$ the Bernoulli polynomials become Bernoulli numbers, $B_\ell(0)=B_\ell$.  For  odd $\ell$, they are 
$B_1=-1/2$ and vanish for $\ell=3,5,\dots$. As a consequence, $Q^{(0)}_{2k+1} =0$ for $y=i\pi$. 
To verify the relation \re{Q-zero}, we  rewrite \re{mom-eq} as
\begin{align}\label{mom-eq1}
 (g\partial_g)^3  Q_\ell = [ 2 g^2\partial_g^2u -4 (u - g\partial_g u)  (g\partial_g) ] Q_\ell  - 16g^2 \partial_g \lr{g \,Q_{\ell+2}}\,.
\end{align}
Solving this equation we 
obtain an iterative relation for $Q_\ell$
\begin{align}\label{Q-iter}
Q_\ell = Q_\ell^{(0)} + \int_0^{g} {dg'\over 2g'} \log^2({g/ g'})\left\{ [ 2 g'{}^2\partial_{g'}^2u -4 (u - g'\partial_{g'} u)  (g'\partial_{g'}) ] Q_\ell  - 16{g'}^2 \partial_{g'} \lr{g' \,Q_{\ell+2}}\right\}
\, ,
\end{align}
where the functions $u$ and $Q_\ell$ on the right-hand side depend on $g'$, and  $Q_\ell^{(0)}$ is given by \re{Q-Ber}. Taking into account that $u=O(g^2)$ 
at weak coupling, we observe that the integral on the right-hand side of \re{Q-iter} is suppressed by a power of $g^2$.
At weak coupling, we substitute $Q_\ell = Q_\ell^{(0)}+ g^2 Q_\ell^{(1)} + \dots$ into \re{Q-iter} and obtain a system of recurrence relations for  
 $Q_\ell^{(k)}$.
   In particular, we find  that $Q_\ell^{(1)}$ is given by a linear combination of $Q_\ell^{(0)}$ and $Q_{\ell+2}^{(0)}$. 
Then, the vanishing of $Q_\ell^{(0)}$ and $Q_{\ell+2}^{(0)}$ for $y=i\pi$ automatically implies that $Q_\ell^{(1)}$ is zero. The same argument can be repeated iteratively order-by-order in $g$. 
As a result, the vanishing of $Q_\ell$ and $Q_{\ell+2}$ for some reference $y$ at any given order in the $g$ implies that $Q_\ell$ should vanish at the next order as well. Since $Q_\ell$ 
vanishes at the lowest order in 't Hooft coupling for $y=i \pi$ and any odd $\ell\ge 3$, the same should be true to any order in $g^2$ leading to \re{Q-zero}. Since $Q_\ell$ is a real function of $y$, it should also vanish for $y=-i\pi$.
This completes the proof.

We would like to stress that the relation \re{Q-zero} is ultimately linked to properties of the cut-off function  \re{chi}. Namely, the 
exponential weight factor $\e^z/(\e^z+1)^2$ that is responsible for the appearance of the Bernoulli polynomial in \re{Q-Ber} arises from $\partial_y \chi(x)$ for $x=4g^2(z+y)^2$.

%%%%%%%%%%%%%%%%%%%%%%%%%%%%%%%%%%%%%%%%%%%%%%%%%%%%%%%%%%%%%%%%%%%%%%%%%%%
\subsection{Potential at weak coupling}
%%%%%%%%%%%%%%%%%%%%%%%%%%%%%%%%%%%%%%%%%%%%%%%%%%%%%%%%%%%%%%%%%%%%%%%%%%%

Deriving the relation \re{Q-zero}, we assumed that $u=O(g^2)$ at weak coupling but its explicit form was irrelevant. In this subsection, we apply the relations \re{Q0-norm} and \re{Q-iter} to find $u$ at weak coupling.

According to \re{Q0-norm}, the zeroth moment $Q_{0}$ is protected from quantum corrections and, therefore, the integral on 
the right-hand side of \re{Q-iter} ought to vanish for $\ell=0$ order-by-order in $g$.  For $\ell=0$ the integral involves the moment $Q_2$.
Recursively applying \re{Q-iter} for $\ell=2,4,\dots$ and replacing  $u=u^{(0)} g^2 + u^{(1)} g^4 + \dots$ we can express $Q_2$ to any given order in coupling in terms of the expansion coefficients $u^{(0)}, u^{(1)},\dots $ and the even moments $Q_2^{(0)}, Q_4^{(0)}, \dots $
Substituting $Q_2$ into \re{Q-iter} and requiring the integral to vanish for $\ell=0$, we find after some algebra
\begin{align} \label{u-sol}
  u^{(0)}= 4 Q_2^{(0)}\,,
\quad
 u^{(1)}= 4 (Q^{(0)}_2)^2-4 Q^{(0)}_4\,,
\quad 
 u^{(2)}=4 (Q^{(0)}_2)^3-6 Q^{(0)}_4Q^{(0)}_2+2 Q^{(0)}_6\,,  \ \dots
\end{align}
According to \re{Q-Ber} and \re{cases}, the moments $Q_{2n}^{(0)}$ are polynomials in $y^2$ of degree $n$. Then, 
we expect from \re{u-sol} that $u^{(n)}$ should be polynomials in $y^2$ of degree  $2n+2$. Replacing $Q_{2n}^{(0)}$ with their explicit expressions \re{Q-Ber}, we find, however, that higher powers of $y^2$ cancel and $u^{(n)}$ become linear functions of $y^2$
\begin{align} \label{u-few}
  u^{(0)}=4 y^2+\frac{4 \pi ^2}{3}\,, 
\qquad
  u^{(1)}=-\frac{16\pi ^2 }{3} y^2-\frac{64 \pi ^4}{45}\,,
\qquad
  u^{(2)}=\frac{128
   \pi ^4  }{15}y^2 +\frac{2048 \pi ^6}{945}, \quad
   \dots
\end{align}
It is straightforward to extend these relations to any loop order.~\footnote{ At any loop order the expansion coefficients take the form
$u^{(k)} = p_k y^2 \pi^{2k}+ r_k \pi^{2k+2}$ with rational $p_k$ and $r_k$.}

The relation \re{u-few} suggests to look for a general expression for $u$ in the form
\begin{align}\label{AB}
 u = y^2 A(g) + B(g)\,.
\end{align}
We deduce from \re{u-few} that the functions $A$ and $B$ have perturbative expansion
\begin{align}\notag\label{AB-weak}
& A= 4 g^2-\frac{16 \pi ^2 g^4}{3}+\frac{128 \pi ^4 g^6}{15}
+O\left(g^{8}\right)\,,
\\
& B= \frac{4 \pi ^2 g^2}{3}-\frac{64 \pi ^4 g^4}{45}+\frac{2048 \pi ^6 g^6}{945}
+O\left(g^{8}\right)   \,.
\end{align}
The functions $A(g)$ and $B(g)$ are not independent. We can apply \re{Q-zero} and \re{mom-eq} to obtain a nontrivial relation 
between them. To this end, we set $\ell=1$ on both sides of \re{mom-eq} and take $y=i \pi$. The expression on the right-hand side of \re{mom-eq} vanishes in virtue of \re{Q-zero}, whereas $Q_1$ on the 
left-hand side can be replaced with \re{Q1}. This leads to 
\begin{align}\label{mom-eq2}
\left[(g\partial_g-2)^3+ 4 (u - g\partial_g u)  (g\partial_g-2) - 2 g^2\partial_g^2u  \right]  \partial_y u = 0\,,\qquad \text{for $y=\pm i\pi$}\,.
\end{align}
Replacing the potential with its expression \re{AB},  $u=-\pi^2 A(g) + B(g)$ for $y=\pm i\pi$, we obtain an equation relating  the two functions $A(g)$ and $B(g)$. We can use \re{AB-weak} to verify that \re{mom-eq2} is satisfied at weak coupling.

We would like to emphasize that the relation \re{AB} follows from the properties of the Bernoulli polynomials. Turning the logic around, we could ask whether there exists another polynomial $Q_{2n}^{(0)}$ of
 the form
\begin{align}\label{Q-ansatz}
Q_{2n}^{(0)} = \int_{-\infty}^\infty dz \, \mu(z) (z+y)^{2n}\,,
\end{align}
such that the expressions for $u^{(\ell)}$ in \re{u-sol} remain linear in $y^2$.  Here, as compared to \re{Q=0}, we replaced $\e^z/(\e^z+1)^2$ with some integration measure $\mu(z)$ satisfying  
$\mu(z)=\mu(-z)$ and $\int_{-\infty}^\infty dz \, \mu(z) =1$. Substituting \re{Q-ansatz} into  \re{u-sol} and requiring $u^{(\ell)}$ to be linear in $y^2$ we find after some algebra
\begin{align}
\mu(z) = c {\e^{z c} \over (\e^{z c}+1)^2} = {c\over 4\cosh^2(cz/2)}\,,
\end{align}
with $c$ arbitrary. The corresponding polynomials coincide (up to multiplication by the factor of $c^{2n}$ and rescaling $y\to y/c$) with the Bernoulli polynomials \re{Q-Ber}. Since the integration 
measure in \re{Q=0} coincides with the cut-off function \re{chi} upon the change of variables \re{change}, this property implies that if the cut-off function $\chi$ were different from \re{chi}, the relation \re{AB} would not hold and the function $u$ 
would  necessarily contain higher powers of $y^2$.

Substitution of \re{AB} into \re{D-u} yields the expected result for the Fredholm determinant \re{goal} with the functions $\Gamma(g)$ and $C(g)$ given by
\begin{align}\label{Gamma-exact}
\Gamma(g)= \int_0^g {d g'\over g'} A(g') \,,\qqqquad
C(g)=- 4 \int_0^g {d g'\over g'} B(g') \,.
\end{align}
Replacing $A$ and $B$ with their expressions at weak coulpling \re{AB-weak}, we reproduce the relation \re{Gamma-weak}.
In the next subsection,  we demonstrate how the relations \re{AB} and \re{mom-eq2} can be used to determine the functions 
\re{Gamma-exact} for an arbitrary value of $g$. 
 
%%%%%%%%%%%%%%%%%%%%%%%%%%%%%%%%%%%%%%%%%%%%%%%%%%%%%%%%%%%%%%%%%%%%%%%%%%%
\subsection{Exact solution}
\label{LargeYSection}
%%%%%%%%%%%%%%%%%%%%%%%%%%%%%%%%%%%%%%%%%%%%%%%%%%%%%%%%%%%%%%%%%%%%%%%%%%%

We show in this subsection that the function $A(g)$ entering \re{AB} can be found from the recurrence relation \re{mom-eq1} at large $y$. 

We recall that $Q_\ell$ is a polynomial in $y$ of degree $\ell$ with a definite parity. 
Replacing in \re{mom-eq1}
\begin{align}\label{f-def}
Q_\ell = y^\ell f_\ell(g)+ O(y^{\ell-2})  
\end{align}
and taking into account \re{AB}, we find that $f_\ell(g)$ satisfies the recurrence relation
\begin{align}\label{eq-f} 
\left[  4 (A - g\partial_g A)  (g\partial_g) -2 g^2\partial_g^2A \right]  f_\ell  = &  - 16g^2 \partial_g \lr{g  f_{\ell+2}}\,.
\end{align}
Notice that the left-hand side of \re{mom-eq1} is suppressed by a power of $1/y^2$ and, therefore, it does not contribute to \re{eq-f}. The boundary conditions for $f_\ell(g)$ 
arise from \re{Q1}
\begin{align} \label{bc}
f_0 =1\,, & \qqqquad f_1 = {1\over 4g^2} A(g)\,.
\end{align}
It is convenient to introduce a generating function for $f_\ell$ 
\begin{align}\label{orig}
f(x,g) 
&=1+ \sum_{\ell=1}^\infty f_\ell(g) x^\ell
=\lim_{y\to\infty}  \int_{-\infty}^\infty  dz  \frac{ {\rm e}^{z}}{ ({\rm e}^{z}+1)^2} {q^2(z)\over 1-x(1+z/y)}
\end{align}
Here in the second relation we used \re{f-def} and replaced $Q_\ell$ with its integral representation \re{mom-def}.
At zero coupling, we replace $q^2(z)=1$ (see Eq.~\re{triv}) to find from \re{orig}
\begin{align}\label{bound}
f(x,0) = {1\over 1-x}\,.
\end{align}
Taking into account \re{eq-f}, we conclude that $f(x,g)$ satisfies a differential equation
\begin{align} 
& \left[  4 (A - g\partial_g A)  (g\partial_g) -2 g^2\partial_g^2A \right]  x^2 f( x,g)  =   - 16g^2 \partial_g \left[{g  (f(x,g)-1-f_1(g) x) }
\right]
\end{align}
where $f_1(g)$ is given by \re{bc}. Solving this equation with the boundary condition \re{bound}, we get 
\begin{align}\label{sq}
f(x,g) =  {1\over g \sqrt{1 -x^2w^2(g)}} \int_0^g {dg'}  \frac{1 +x w^2(g')}{  \sqrt{1-x^2 w^2(g')}} \,,
\end{align}
where we introduced a shorthand notation for
\begin{align}\label{v-A}
w^2(g) = \partial_g\left(A(g)\over 4g\right)\,.
\end{align}
At weak coupling, we apply \re{AB-weak} to get $w^2(g)=1-2 \pi ^2 g^2+O\left(g^4\right)$. It is easy then to check that for $g\to 0$, the relation \re{sq} reduces to \re{bound}.
 
To determine the function $w(g)$, we match analytical properties of expressions on the right-hand sides of \re{orig} and \re{sq}. According to \re{sq}, the 
function $f(x,g)$ has singularities at $x=\pm 1/w(g)$ and 
$x=\pm 1/w(g')$ generated by the two square roots on the right-hand side of \re{sq}. At weak coupling, for $w^2(g)=1 + O(g^2)$, they are located in the vicinity of $x=\pm 1$. 
Let us now examine Eq.\ \re{orig}. The integrand in \re{orig} has a pole at $x=1/(1+z/y)$ and, as a consequence, the function $f(x,g)$ has a discontinuity in $x$. Since 
the integration measure in \re{orig} suppresses the contribution from large $z$, the cut is located at $x=1+O(1/y)$. Thus, the function $f(x,g)$ has to be 
regular in the vicinity of $x=-1$.

 Imposing this condition on  \re{sq}, we obtain a nontrivial relation for the function $w(g)$. Indeed, requiring the expression on the 
right-hand side of \re{sq} to be finite for $x=-1/w(g)$, we find that the integral in \re{sq} has to vanish
\begin{align}\label{int-eq}
\int_0^g {dg'}  \frac{1 - w^2(g')/w(g)}{  \sqrt{1-w^2(g')/w^2(g)}}  = 0\,.
\end{align}
It is straightforward to solve this relation  at weak coupling. Replacing $w(g) = \sum_{k\ge 0} (-1)^k w_k g^{2k}$ and equating 
to zero the coefficients in front of powers of $g^2$ on the left-hand side of \re{int-eq}, we get
\begin{align}
w_0=1\,,\qqquad w_2=\frac{5 w_1^2}{6}\,,\qqquad
w_3= \frac{61 w_1^3}{90}\,,\qqquad
w_4= \frac{277 w_1^4}{504}\,,\qquad
\dots
\end{align}
with $w_1$ being arbitrary. With some guesswork we can resum the series in $g^2$ to get
\begin{align}\label{v-sol}
w(g)={1\over \cosh(g\sqrt{2w_1})} \,.
\end{align}
Substituting this expression into \re{int-eq} we verify that the relation \re{int-eq} is satisfied for arbitrary $g$ and $w_1$. The ambiguity in $w_1$ is due to the fact that the relation 
\re{int-eq} is invariant under rescaling of the coupling. To fix $w_1$, it is sufficient to compare \re{v-sol} with its expansion at weak coupling, $w^2(g)=1-2 \pi ^2 g^2+O\left(g^4\right)$ (see Eqs.~\re{v-A} and \re{AB-weak}). In this way, we get $w_1=2\pi^2$. 

We are now ready to determine the functions $A(g)$ and $B(g)$ defined in \re{AB}. Replacing $w(g)$ in \re{v-A} with its expression \re{v-sol} (for $w_1=2\pi^2$), we find
\begin{align}\label{A-res}
A(g)  =2\pi g\tanh(2\pi g)/\pi^2\,.
\end{align}
At weak coupling this relation agrees with \re{AB-weak}. 
It is interesting to note that the function $A(g)$ obeys a non-linear differential equation
\begin{align}
4-\partial_g ( A(g)/g) - \pi^2 \lr{A(g)/g}^2=0\,.
\end{align}
Together with the definition \re{v-A}, it can be used to prove the condition \re{int-eq}. Indeed, changing the integration variable as $z=w^2(g')$, we can recast the integral in \re{int-eq} 
into the form
\begin{align}
\int_1^{z_0} {dz\, (\sqrt{z_0} -z)\over z\sqrt{ (z_0-z)(z-1)}} = \oint {dz\over 2  i} { \sqrt{z_0} -z \over z\sqrt{ (z-z_0)(z-1)}} \,,
\end{align}
where $z_0=w^2(g)$ and the integration goes along the contour that encircles the cut $[1,z_0]$ in a clockwise direction. Blowing up the integration contour, we find by means of the
Cauchy theorem, that the integral is given by the sum of residues at the poles at $z=0$ and $z=\infty$. This sum vanishes and we recover the right-hand side of \re{int-eq}.

To determine the function $B(g)$, we apply the relation \re{mom-eq2}. For $y=\pm i\pi$ we replace $u=-\pi^2 A(g) + B(g)$ there and derive a differential equation for $B(g)$. 
Its solution takes the form 
\begin{align}\label{B-res}
B(g)= \pi  g \coth (4 \pi  g)-\frac{1}{4}\,.
\end{align}
We verify that its expansion at weak coupling is in agreement with \re{AB-weak}.

Combining together the relations \re{AB}, \re{A-res} and  \re{B-res}, we finally obtain the exact expression for the potential 
\begin{align}
u =  2\pi g\tanh(2\pi g){y^2\over \pi^2} + \pi  g \coth (4 \pi  g)-\frac{1}{4}\,.
\end{align}
As expected, $u$ is a linear function of $y^2$ with  coefficients being nontrivial functions of the coupling constant. 

To conclude our analysis, we apply \re{D-u} and compute the 
logarithm of the Fredholm determinant, or equivalently  the 
logarithm of the null octagon
\begin{align}
\log \mathbb O = d(y,g)=-{y^2\over 2 \pi ^2}  {\log (\cosh (2 \pi  g))} -\frac{1}{8} \log \left({\sinh (4 \pi  g)\over 4 \pi  g}\right)\,.
\end{align}
Comparing this relation with  \re{goal}, we arrive at the exact expressions for the anomalous dimensions \re{Gam}.
 
%%%%%%%%%%%%%%%%%%%%%%%%%%%%%%%%%%%%%%%%%%%%%%%%%%%%%%%%%%%%%%%%%%%%%%%%%%%
\section{Conclusions}
\label{ConclusionsSection}
%%%%%%%%%%%%%%%%%%%%%%%%%%%%%%%%%%%%%%%%%%%%%%%%%%%%%%%%%%%%%%%%%%%%%%%%%%%

In this paper, we studied the so-called simplest correlation function of four infinitely heavy half-BPS operators in planar $\mathcal N=4$ SYM 
introduced in Refs.~\cite{Coronado:2018ypq,Coronado:2018cxj} in the limit when the operators are null separated in a sequential manner. 
 Our analysis heavily relied on the factorization of this correlation function into the product of two octagons and on the
determinant representation for the latter established in Refs.~\cite{Kostov:2019stn,Kostov:2019auq}. We argued that the null octagon is given by a 
Fredholm determinant of a certain integral operator with the integrable Bessel kernel modified by Fermi-Dirac like distribution. In this representation,
the 't Hooft coupling constant plays the role of the temperature and the cross ratio defines the chemical potential. 

The integral operator defining the null octagon has a striking similarity to those previously encountered in the study of two-point correlation functions in 
exactly solvable models at finite temperature and of level spacing distributions for random matrices. In application to these two problems,
a poweful method of differential equations has been developed allowing one to compute various physical quantities starting from their determinant 
representation~\cite{Its:1990,Korepin:1993kvr,Tracy:1993xj}. 
We generalized this approach to the Fredholm determinant at hand and successfully derived a system of nonlinear differential equations for the null 
octagon as a function of the 't Hooft coupling.
Solving these equations we found a closed-form expression for the null octagon, thus, justifying the title of the 
paper.

There are several avenues which can be explored further. 

Having determined the exact  functional form of the null octagon, Eqs.~\re{LC} and \re{Gam}, we can examine its strong coupling expansion. We find from \re{Gam} 
that the function $\Gamma (g)/\pi^2$ accompanying the $(-y^2/2)$ term in \re{LC} admits the following expansion when $g^2$ is large
\begin{align} \label{str} 
 \Gamma(g) /\pi^2 {}& = \frac{2 g }{ \pi} -  {\log 2\over \pi^2} - {1\over \pi^2} \sum_{n \geq 1} \frac{(-1)^n}{n} \e^{-4\pi g n}
\, .
\end{align}
From the point of view of the dual string theory description, the first term on the right-hand side comes from the classical action of the open string with four boundaries 
defined by geodesics in AdS${}_5$, the second term describes quadratic fluctuations of the world-sheet and the last one  contains an infinite sum of nonperturbative 
exponentialy-small corrections. Notice that the expansion does not contain higher order corrections in  $1/g$. This suggests that the path integral describing the open string partition function localizes at the classical configuration.  
This point deserves further study. 

The relation \re{str} should be compared with the strong coupling expansion of the cusp anomalous dimension
\begin{align}
 \Gamma_{\rm cusp}(g) = 2g -{3\log 2\over 2\pi}  + O(1/g) + O(g^{1/2} \e^{-2\pi g})\,.
\end{align}
As was mentioned in the Introduction, $ \Gamma_{\rm cusp}(g)$ governs the leading light-like asymptotics
of the four-point correlation function of shortest (length $K=2$) half-BPS operators and, therefore, it can be thought of as a counter-partner of \re{str}.
At weak coupling, $2\Gamma(g) /\pi^2$ and $\Gamma_{\rm cusp}(g)$ coincide at one loop and differ starting from two loops. At strong coupling,  the two functions
differ by the factor of $2/\pi$ already at leading order.

Having the machinery of differential equations at our disposal, we can generalize consideration in two different directions.  
As was shown in  Refs.~\cite{Kostov:2019stn,Kostov:2019auq}, the octagon admits a determinant representation for generic values of cross ratios \re{cross}. Although it simplifies significantly in the light-like limit, $z\to 0_-$ and $\bar z\to -\infty$,
the method of differential equation should be applicable to studying the octagon for generic values of $z$ and $\bar z$. In this paper, we analyzed the simplest octagon with zero internal
bridges. It would be interesting to extend the consideration to more complicated octagons with $\ell > 0$ internal bridges defining another family of exact correlation functions, dubbed asymptotic \cite{Coronado:2018ypq}. 

Remarkable simplicity of the functions \re{Gam} calls for explanations. It also suggests that there should exist another, simpler way of deriving \re{Gam}. Indeed, 
in the course of our analysis, we uncovered two different integral equations for the function $\Gamma (g)$. The first one is closely related to the
Fredholm determinant representation of the octagon. We found that $\Gamma (g)$ arises from the relation
\begin{align}
\label{BMNgamma}
\Gamma (g) = 2 \pi^2 \int_0^\infty \frac{dt}{t} \left( \frac{1}{1 + {\rm e}^t} \right)^{\prime\prime} G (t,t)
\, ,
\end{align}
where $G(t,t)$ is the limiting value of the function $G (t,t')$ that obeys a Schwinger-Dyson type equation,
\begin{align}\label{SD}
G (t_1, t_2) - 2 \int_0^\infty \frac{dt_3}{t_3} \frac{G (t_1, t_3)}{1 + {\rm e}^{t_3}}  H (2 g t_3, 2 g t_2)
=
H (2 g t_1, 2 g t_2)
\, .
\end{align}
Here the function $H (2 g t_1, 2 g t_2)$ is defined in \re{H-def}.  It was introduced by the hexagonalization procedure and describes excitations propagating on 
top of the BMN vacuum. At weak coupling, it is easy to solve \re{SD} by iterations and obtain a representation for $G(t_1,t_2)$ as iterated integrals involving  the kernel $H (2 g t_1, 2 g t_2)$.

The second equation for $\Gamma (g)$ takes the form
\begin{align}\label{GKP1}
\Gamma (g) = g \pi^2 \int_0^\infty \frac{dt}{t} J_1 (2 g t) \gamma (2 g t)
\, ,
\end{align}
where $J_1 (2 g t)$ is the Bessel function and the function $\gamma (- t) = - \gamma (t)$ obeys an infinite system of equations
\begin{align}\label{GKP2}
\int_0^\infty \frac{dt}{t} J_{2n-1} (2 g t) \coth(t/2) \gamma (2 g t) = 2 g \delta_{n,1}
\, ,
\end{align}
with $n=1,2,\dots$. 
This relation  has a striking similarity  with the flux-tube equation for the GKP vacuum. However, in distinction to the latter case, it only involves an odd function of the spectral 
parameter $\gamma(t)$ and Bessel functions of odd order. The equations \re{GKP2} can be solved at weak coupling by iterations and at strong coupling by converting them into
singular integral equations via a Fourier transform. In both cases, we find that the resulting expression for $\Gamma(g)$ is in agreement with \re{Gam}. The origin of the equations 
\re{GKP1} and \re{GKP2} remains obscure. 

The existence of the two complimentary descriptions, Eqs.~\re{SD} and \re{GKP2}, for the same function $\Gamma(g)$ strongly suggests to search for a functional transformation 
between them. In GKP representation, $\Gamma(g)$ is encoded in a single Bessel moment \re{GKP1}, whereas BMN representation \re{BMNgamma} involves an infinite number of 
them, once one recasts the formula making use of a completeness conditions for Bessel functions. We leave this interesting problem for future studies.

\section*{Acknowledgments}

We would like to thank Ivan Kostov and  Didina Serban for collaboration at the early stage of this work. We are also grateful to 
Philippe Di Francesco, Bertrand Eynard, Ivan Kostov and  Didina Serban for useful discussions, and Pedro Vieira for comments on the draft.
The research of A.B.\ was supported 
by the U.S. National Science Foundation under the grant PHY-1713125 and of G.K.\ by the French National Agency for Research grant ANR-17-CE31-0001-01.

\appendix

\section{Differential equations}    
    
In this Appendix, we present a derivation of the differential equations for the auxiliary functions and potentials defined in \re{PQ} and \re{u-def}, respectively.
The analysis goes along the same lines as in Ref.~\cite{Tracy:1993xj}.

We start with the first relation in \re{u-def} and differentiate its both sides with respect to $\alpha=\{y,g\}$. Since the dependence on $\alpha$ resides in $\bbchi$ we get
\begin{align}\notag
\partial_\alpha u  &= \vev{\phi|\partial_\alpha \bbchi  {1\over 1-\mathbb{K} \bbchi}|\phi}+
 \vev{\phi|\chi  {1\over 1-\mathbb{K} \bbchi}\mathbb{K} \partial_\alpha\chi{1\over 1-\mathbb{K} \bbchi}|\phi}
 \\
 & = \vev{\phi|  {1\over 1-\bbchi \mathbb{K}}  \partial_\alpha\bbchi {1\over 1-\mathbb{K}\bbchi }|\phi} = \int dx\, Q(x)Q(x)  \partial_\alpha \chi(x)\,.
 \end{align}
 Here in the last relation we replaced $\bbchi=\int_0^\infty \ket{x}\chi(x)\bra{x}$ and took into account \re{PQ}.
In the similar manner,
\begin{align}  \notag
\partial_\alpha v  &= \vev{\phi|\partial_\alpha \bbchi  {1\over 1-\mathbb{K} \bbchi}|\psi}+
 \vev{\phi|\bbchi  {1\over 1-\mathbb{K} \bbchi}\mathbb{K} \partial_\alpha\bbchi{1\over 1-\mathbb{K} \bbchi}|\psi}
 \\
 & = \vev{\phi|  {1\over 1-\bbchi \mathbb{K}}  \partial_\alpha\bbchi {1\over 1-\mathbb{K}\bbchi }|\psi} = \int dx\, Q(x)P(x)  \partial_\alpha \chi(x)\,.
\end{align}
Then, we evaluate the logarithmic derivative of the function $Q(x)$ defined in \re{PQ}
 \begin{align}\notag\label{dQ}
x Q'(x) &=  \bra{x} \mathbbm{x} \partial_\mathbbm{x} {1\over 1- \mathbb{K}\bbchi } \ket{\phi} = \bra{x}  {1\over 1- \mathbb{K}\bbchi }  \mathbbm{x}  \partial_ \mathbbm{x}  \ket{\phi} +  \bra{x}[ \mathbbm{x}  \partial_ \mathbbm{x} , {1\over 1- \mathbb{K}\bbchi } ]\ket{\phi} 
\\\notag
& = \bra{x} {1\over 1- \mathbb{K}\bbchi }\ket{\psi} + \frac14 \bra{x} {1\over 1- \mathbb{K}\bbchi } \ket{\phi}\bra{\phi}\bbchi  {1\over 1- \mathbb{K}\bbchi }\ket{\phi} +\bra{x} {1\over 1- \mathbb{K}\bbchi } \mathbb{K} \mathbbm{x} \partial_\mathbbm{x} \bbchi  {1\over 1- \mathbb{K}\bbchi }\ket{\phi}
\\
&= P(x) + \frac14 Q(x) u +\bra{x} {1\over 1- \mathbb{K}\bbchi } \mathbb{K} \mathbbm{x} \partial_\mathbbm{x} \bbchi  {1\over 1- \mathbb{K}\bbchi }\ket{\phi}\,.
\end{align}
Here in the second relation, we used the operator identity
\begin{align}\notag \label{com} 
[  \mathbbm{x} \partial_{  \mathbbm{x}} , {1\over 1- \mathbb{K}\bbchi }] & = {1\over 1- \mathbb{K}\bbchi }[  \mathbbm{x} \partial_{  \mathbbm{x}} ,  \mathbb{K}\bbchi ]  {1\over 1- \mathbb{K}\bbchi }
 \\ & 
= {1\over 1- \mathbb{K}\bbchi } \frac14  \ket{\phi}\bra{\phi}\bbchi   {1\over 1- \mathbb{K}\bbchi }+{1\over 1- \mathbb{K}\bbchi } \mathbb{K} \mathbbm{x} \partial_\mathbbm{x} \bbchi  {1\over 1- \mathbb{K}\bbchi }\,.
\end{align}
To evaluate the last term in \re{dQ}, we use the property \re{chi-rel} of the  function $\chi$ and replace it with
\begin{align}\notag
- \frac12 \bra{x} {1\over 1- \mathbb{K}\bbchi } \mathbb{K} g \partial_g \bbchi  {1\over 1- \mathbb{K}\bbchi }\ket{\phi} 
 = - \frac12 g \partial_g  \bra{x} {1\over 1- \mathbb{K}\bbchi } \ket{\phi} =-\frac12 g\partial_g Q(x)\,.
\end{align}
In this way, we arrive at
\begin{align}
x Q'(x) = P(x) + \frac14 Q(x) u -\frac12 g\partial_g Q(x)\,.
\end{align}
Analogously,
\begin{align} \notag
x P'(x) &= \bra{x}  {1\over 1- \mathbb{K}\bbchi } \mathbbm{x} \partial_\mathbbm{x} \ket{\psi}  + \frac14 Q(x) v +\bra{x} {1\over 1- \mathbb{K}\bbchi } \mathbb{K}    \mathbbm{x} \partial_\mathbbm{x} \log \bbchi  {1\over 1- K \bbchi}  \ket{\psi}
\\\notag
&=-{x\over 4} Q(x) + \frac12 Q(x) v-\frac14 P(x) u + \bra{x} {1\over 1- \mathbb{K}\bbchi } \mathbb{K}    \mathbbm{x} \partial_\mathbbm{x} \log \bbchi  {1\over 1- K \bbchi}  \ket{\psi}
\\
&=-{x\over 4} Q(x) + \frac12 Q(x) v-\frac14 P(x) u -\frac12 g\partial_g P(x)
\, .
\end{align}    
Here in the first relation, we used the property of the Bessel function $x \partial_x \psi(x) = -x\phi(x)/4$.
 
To prove the relation \re{uv-rel},  we consider the following quantity
\begin{align}\notag\label{uv-chain}
\partial_\alpha \lr{v+\frac18 u^2} &= \int dx \, Q(x)\left[ P(x) + \frac14 u \,Q(x) \right]  \partial_\alpha\chi (x)
  = \int dx \, \partial_\alpha\chi (x) \left( \frac12 x \partial_x  +\frac14 g\partial_g  \right) Q^2(x)   
\\
&=  \lr{-\frac12   +\frac14 g\partial_g }\int dx \, Q(x) Q(x)  \partial_\alpha\chi (x)=  \lr{-\frac12 +\frac14 g\partial_g } \partial_\alpha  u\,.
\end{align}
Here in the first line, we applied \re{der1} and  took into account the first relation in \re{der2}. In the second line, we integrated by parts
one of the terms and used the relation \re{chi-rel}. It follows
from \re{uv-chain} that $v+\frac18 u^2+\lr{\frac12 -\frac14 g\partial_g } u$ is a constant independent on $g$ and $y$. To fix its values, it suffices to consider the limit $g\to0$ with $y$ kept fixed. 
According to \re{step}, the integral operator $\mathbb K$ acts  in this limit  on the interval $[0,s]$ that shrinks into a point as $g\to 0$. As a result, $u$ and $v$ vanish in this limit and the 
aforementioned constant should be zero.    
    
According to \re{R-def},
the function $R(x,y)\chi(y)$ is  a kernel of the operator $\lr{1- \mathbb{K} \bbchi}^{-1}  \mathbb{K}\bbchi$.  
As a consequence, it obeys the equation
\begin{align}\notag
(x\partial_x + y \partial_y +1) & \lr{R(x,y)\chi(y)} = \vev{x| [  \mathbbm{x} \partial_ {\mathbbm{x}} , { \mathbb{K}\bbchi \over 1- \mathbb{K}\bbchi }] |y}= \vev{x| [  \mathbbm{x} \partial_ {\mathbbm{x}}  , {1\over 1- \mathbb{K}\bbchi }] |y}
\\\notag
&= \frac14 \bra{x} {1\over 1- \mathbb{K}\bbchi } \ket{ \phi}\bra{\phi}\bbchi  {1\over 1- \mathbb{K}\bbchi } \ket{y} +
  \vev{x|{1\over 1- \mathbb{K}\bbchi } \mathbb{K} \mathbbm{x} \partial_ {\mathbbm{x}}  \bbchi_s  {1\over 1- \mathbb{K}\bbchi } |y} 
  \\ &
= \frac14 Q(x) Q(y)\chi (y) -\frac12 g\partial_g  \lr{R(x,y)\chi(y)}
\, ,
\end{align}
where in the first line we used \re{com} and applied \re{chi-rel} in the second line.
For $y\to x$  this relation simplifies to
\begin{align}\notag
&\lr{1+x \partial_x} \lr{R(x,x)\chi(x) }+\frac12 g\partial_g  \lr{R(x,x) \chi(x)}   
\\
& = \chi(x) \left( 1+x \partial_x+ \frac12 g\partial_g  \right) R(x,x)   = \frac14 Q^2(x)\chi (x)
\, ,
\end{align}    
where we used \re{chi-rel}.

\bibliographystyle{JHEP} 

\bibliography{papers}

\end{document}